\documentclass{emulateapj}
\usepackage{graphicx}

\newcommand {\kms}    {km~s$^{-1}$}

\slugcomment {To appear in the Astrophysical Journal}
\shorttitle{Ejecta Flickering, Ablation, and Fragmentation}

\begin{document}

\title{Ejecta Knot Flickering, Mass Ablation, and Fragmentation in Cassiopeia A{\altaffilmark{1}} }

\author{Robert A.\ Fesen, Jordan Zastrow, Molly C.\ Hammell}
\affil{6127 Wilder Lab, Department of Physics \& Astronomy, Dartmouth
  College, Hanover, NH 03755}
\and
\author{J.\ Michael Shull \& Devin W. Silvia}
\affil{CASA, Department of Astrophysical and Planetary Science, University of Colorado,
   Boulder, CO 80309}

\altaffiltext{1} {Based on observations with the NASA/ESA Hubble Space Telescope, 
obtained at the Space Telescope Science Institute, 
which is operated by the Association of Universities for Research in Astronomy, Inc.}

\begin{abstract}

Ejecta knot flickering, ablation tails, and fragmentation are expected
signatures associated with the gradual dissolution of high-velocity supernova
(SN) ejecta caused by their passage through an inhomogeneous circumstellar or
interstellar medium.  Such phenomena mark the initial stages of the gradual
merger of SN ejecta with and the enrichment of the surrounding interstellar
medium. Here we report on an investigation of this process through changes in
the optical flux and morphology of several high-velocity ejecta knots located
in the outskirts of the young core-collapse SN remnant Cassiopeia A using {\sl
Hubble Space Telescope} images.  Examination of WFPC2 F675W and  combined ACS
F625W + F775W images taken between June 1999 and December 2004 of several dozen
debris fragments in the remnant's northeast ejecta stream and along the
remnant's eastern limb reveal substantial emission variations (`flickering')
over time scales as short as nine months. Such widespread and rapid variability
indicates knot scale lengths $\simeq 10^{15}$ cm and a highly inhomogeneous
surrounding medium. We also identify a small percentage of ejecta knots located
all around the remnant's outer periphery which show trailing emissions
typically $0\farcs2 - 0\farcs7$ in length aligned along the knot's direction of
motion suggestive of knot ablation tails. We discuss the nature of these
trailing emissions as they pertain to ablation cooling, knot disruption and
fragmentation, and draw comparisons to the emission ``strings'' seen in $\eta$
Car.  Finally, we identify several tight clusters of small ejecta knots which
resemble models of shock induced fragmentation of larger SN ejecta knots caused
by a high-velocity interaction with a lower density ambient medium.

\end{abstract}


\section{Introduction}

Supernova (SN) ejecta can enrich a galaxy's interstellar medium with
nucleosynthesis products generated during the progenitor's main sequence
lifetime, post-main sequence evolution, and the explosive burning that occurs
during the supernova outburst itself.  In this way, SNe influence a galaxy's
chemistry by steadily increasing the metallicity of subsequent stellar
populations (see \citealt{MG86}, \citealt{Pagel97}, \citealt{Scalo04},
\citealt{Matt06} and references therein).  

Expanding shells of stellar debris can undergo instabilities leading to the
formation of ejecta clumps \citep{Nag88,Mueller91,Kif03,Hammer2010}.
Indeed, there is considerable observational evidence for ejecta clumping
especially in core-collapse SNe
(e.g., \citealt{Fil89,Spy91,Spy94,Fassia98,Math00,Elm04}).

Unfortunately, there are few direct observational constraints regarding the
timescales for the fragmentation and dissolution of ejecta clumps as they move
through and merge with the progenitor's circumstellar medium (CSM) and local
interstellar medium (ISM).  In contrast to this scarcity of direct
observations, the interaction of SN ejecta fragments with local CSM and ISM in
supernova remnants (SNR) has been the subject of many theoretical studies
(e.g., \citealt{Chevalier75,Hamilton85,Jones94,Anderson94,Cid96,B02,WC02}).
Several related studies have also discussed the properties of the interstellar
``bullets'' seen in Herbig-Haro (HH) objects and planetary nebulae
\citep{NS79,Pol04a,Raga07} and there are numerous investigations addressing the
general case of a shock wave interacting with interstellar clouds (see
\citealt{Woodward76,SN92,Klein94,MacLow94,Pittard09,Pittard10} and references
therein).

Young and relatively nearby Galactic SNRs offer the possibility of relatively
high resolution investigations of such SN ejecta ISM/CSM interactions.  With
the possible exception of the large ejecta clumps seen in the Vela remnant
\citep{Aschen95,Strom95,Tsun99,Kat06}, the Galactic SNR exhibiting the best
example of SN ejecta clumps is the young core-collapse remnant Cassiopeia A
(Cas A).  With a current estimated age of around 330 years
\citep{Thor01,Fes06b} Cas A is believed to be the remains of a Type IIb
supernova event probably from a red supergiant progenitor with an initial mass
in the range of 15--25 M$_{\odot}$ which might have lost much its hydrogen
envelope due to a binary interaction \citep{Young06,Krause08}. 

Cas A consists of an optical, infrared, and X-ray bright $4'$ diameter
($\simeq$4 pc at 3.4 kpc; \citealt{Reed95}) emission ring of reverse 
shock-heated SN debris rich in O, Si, S, Ar, Ca and Fe
\citep{ck78,ck79,Douvion99,Hughes00,Willingale02,Hwang03}.  The remnant's
optically visible debris appears as condensations and filaments with typical
scale lengths of $0\farcs2 - 1\farcs0$ ($1 - 5 \times 10^{16}$ cm)

Early broadband photographic images of Cas A taken over several decades
starting in the 1950's revealed substantial changes in the brightness and
appearance of the remnant's bright main shell of reverse shocked ejecta 
\citep{kvdb76,vdbk85}. These changes have significantly altered the remnant's
overall optical appearance from that seen in 1951 when the remnant was first
optically imaged (see \citealt{vdbD70}).  Changes in the brightness and
appearance of ejecta clumps are fairly gradual, with brightening and fading
e-folding time scales of around 25 years \citep{kvdb76}.

Outside the remnant's main shell of X-ray, optical, and infrared emitting SN
debris lie many hundreds of small ($\lesssim 0\farcs5$) optical emission knots
with expansion velocities up to 14,000 km s$^{-1}$ \citep{Fesen01,Fes06b,HF08}.
While such outlying ejecta are found around much of Cas~A's limb, the
majority lie in the nearly opposing NE and SW ejecta streams or `jets'
producing an apparent bipolar asymmetry
\citep{Fesen01,Hwang04,Krause05,Fes06a}.

To date, no investigation has been made into possible optical flux changes in
any of the many hundreds of small ejecta clumps or `knots' lying outside the
remnant's main shell.  With locations at or ahead of the remnant's forward
shock front \citep{Fes06a,HF08}, they comprise the vanguard of remnant's SN
ejecta and are thus the first to dissolve into and enrich the local ISM. They
are not visible due to reverse shock heating as in the main ejecta shell, but
are instead shock-heated due to their high speed passage through the local
circumstellar and interstellar medium.

Here we report results of an imaging survey of some of these outlying debris
knots which reveal significant brightness fluctuations over times scales as
short as nine months. In addition, a small percentage of knots exhibit
trailing emission suggestive of mass stripping and knot disruption due to
their high-velocity interaction with local CSM and ISM.  We also identify
several tight clusters of ejecta knots with a morphology resembling recent model
results for the fragmentation of much larger ejecta knots. The
observations and knot flux measurement procedures are described in $\S$2 with
the observational results along with some knot disruption models presented in
$\S$3 and discussed in $\S$4.

\section{Observations}

Broadband images of the Cas~A remnant were obtained using two different cameras
on board the {\it Hubble Space Telescope} ({\sl HST}).  Although the primary
focus of these images was Cas A's bright main emission shell of supernova
debris,  some of these images cover parts of the remnant's outer periphery,
thereby detecting dozens of high-speed outlying ejecta knots.
 
Four 500 s exposures of the remnant's easternmost limb were taken on 2000
January 23 and 2002 January 16 using the Wide Field Planetary Camera 2 (WFPC2)
with the F675W filter. WFPC2 images have an image scale of $0\farcs1$
pixel$^{-1}$ which under-sample {\sl HST's} $0\farcs046$ angular resolution.
The F675W filter has a bandpass of $6000 - 7600$ \AA \ and is sensitive to line
emissions of [O~I] $\lambda\lambda$6300, 6364, [N~II] $\lambda\lambda$6548, 6583,
[S~II] $\lambda\lambda$6716, 6731, [Ar III] $\lambda$7136, and  [O~II]
$\lambda\lambda$7319, 7330.  

Additional WFPC2 images of parts of the remnant's northeast jet of ejecta were
obtained in June 1999.  These 1999 jet images
consisted of four 600 s.  Further
descriptions of the WFPC2 data and their reduction can be found in
\citet{Fesenetal01} and \citet{Morse04}.

{\sl HST} images covering the entire remnant, including all previously known
outlying ejecta knots, were obtained on 2004 March 4--6 and 2004 December 4--5
using the Wide Field Channel (WFC) of the Advanced Camera for Surveys (ACS;
\citealt{Ford98,Pavlovsky04}) on board {\sl HST}.  The ACS/WFC consists of two
$2048 \times 4096$ CCDs providing a field of view $202'' \times 202''$ with an
average pixel size of $0\farcs05$.  Dithered images were taken in each of the
four ACS/WFC Sloan Digital Sky Survey (SDSS) filters, namely F450W, F625W,
F775W, and F850LP (i.e., SDSS g,r,i, and z filters) to permit cosmic ray removal,
coverage of the $2\farcs5$ interchip gap, and to minimize saturation effects of
bright stars in the target fields.  Total integration times in these filters were
2000 s, 2400 s, 2000 s, and 2000~s, respectively.  Further descriptions of
these data and their reduction can be found in \citet{Fes06a,Fes06b} and
\citet{HF08}.

\begin{deluxetable*}{ccrrrrcc}
\tablecolumns{8}
\tablecaption{Observed Fluxes of Selected NE Jet Ejecta Knots }
\tablewidth{0pt}
\tablehead{
\colhead{Knot ID}  &
\colhead{Catalog\tablenotemark{a}} &
\multicolumn{2}{c}{\underline{~~~~~~~~~F775W\tablenotemark{b} ~~~~~~~~~~}} &
\multicolumn{2}{c}{\underline{~~~~~~~~~F850LP\tablenotemark{b}~~~~~~~~~}} & 
\multicolumn{2}{c}{\underline{~~Dec 2004/Mar 2004~~}} \\
\colhead{(Fig.\ 3)} &
\colhead{Number} &
\colhead{Mar 2004} &
\colhead{Dec 2004} &
\colhead{Mar 2004} &
\colhead{Dec 2004} &
\colhead{F775W}    &
\colhead{F850LP}  }
\startdata
1 & 377 & $6.4 \pm0.4$ & $20.6 \pm1.3$ &  $15.2 \pm0.4$  & $57.8 \pm3.1$ & $3.2 \pm0.4$ & $3.8 \pm0.3$ \\
2 & 381 & $1.9 \pm0.3$ & $11.7 \pm0.7$ &  $5.9  \pm0.6$  & $36.8 \pm2.2$ & $6.2 \pm2.3$ & $6.2 \pm1.2$ \\
3 & --- & $1.3 \pm0.3$ & $6.6  \pm0.5$ &  $1.6  \pm0.3$  & $7.3  \pm0.4$ & $5.1 \pm2.0$ & $4.6 \pm1.3$ \\
4 & 367 & $2.5 \pm0.4$ & $9.9  \pm0.5$ &  $4.3  \pm0.4$  & $12.7 \pm1.1$ & $4.0 \pm1.0$ & $3.0 \pm0.6$ \\
5 & --- & $2.5 \pm0.4$ & $19.3 \pm1.4$ &  $5.5  \pm0.5$  & $35.7 \pm2.7$ & $7.7 \pm2.2$ & $6.5 \pm1.2$ \\
6 & 217 & $5.3 \pm0.3$ & $40.8 \pm1.9$ &  $2.3  \pm0.2$  & $11.0 \pm0.5$ & $7.7 \pm0.9$ & $4.8 \pm0.7$ \\
7 & 180 & $12.8\pm0.4$ & $5.9  \pm0.5$ &  $2.7  \pm0.5$  & $1.7  \pm0.4$ & $0.5 \pm0.1$ & $0.6 \pm0.1$ \\
8 & 581 & $4.0 \pm0.3$ & $13.5 \pm0.4$ &  $11.0 \pm0.7$  & $41.3 \pm2.5$ & $3.4 \pm0.4$ & $3.8 \pm0.5$ \\
9 & 597 & $3.7 \pm0.3$ & $20.3 \pm0.9$ &  $12.0 \pm0.7$  & $63.6 \pm2.6$ & $5.5 \pm0.7$ & $5.3 \pm0.9$ \\
10& 605 & $7.2 \pm0.5$ & $3.4  \pm0.3$ &  $18.4 \pm0.6$  & $9.6  \pm0.9$ & $0.5 \pm0.1$ & $0.5 \pm0.1$ \\
11& 713 & $3.5 \pm0.5$ & $1.9  \pm0.4$ &  $7.1  \pm0.7$  & $3.5  \pm0.3$ & $0.5 \pm0.1$ & $0.5 \pm0.1$ \\
12& 782 & $1.6 \pm0.3$ & $4.0  \pm0.3$ &  $3.3  \pm0.6$  & $8.7  \pm0.5$ & $2.5 \pm0.8$ & $2.6 \pm0.3$ \\
13& 992 & $5.9 \pm0.4$ & $16.3 \pm0.7$ &  $2.6  \pm0.8$  & $7.1  \pm0.4$ & $2.8 \pm0.3$ & $2.7 \pm0.3$ \\
14& 914 & $8.0 \pm0.5$ & $4.1  \pm0.4$ &  $8.2  \pm0.5$  & $4.4  \pm0.3$ & $0.5 \pm0.1$ & $0.5 \pm0.1$ \\
15& 837 & $6.6 \pm0.5$ & $3.7  \pm0.4$ &  $4.8  \pm0.6$  & $3.1  \pm0.3$ & $0.6 \pm0.1$ & $0.6 \pm0.1$ \\
\enddata
\tablenotetext{a}{From the catalog of outer ejecta knots in the Cas A remnant \citep{HF08}.}
\tablenotetext{b}{Fluxes are in units of 10$^{-17}$ erg cm$^{-2}$ s$^{-1}$}
\end{deluxetable*}

\begin{deluxetable*}{ccrrrr}
\tablecolumns{6}
\tablecaption{Observed Fluxes of Selected Eastern Outer Ejecta Knots }
\tablewidth{0pt}
\tablehead{
\colhead{Knot ID} &
\colhead{Catalog\tablenotemark{a}} &
\colhead{Jan 2000\tablenotemark{b}} &
\colhead{Jan 2002\tablenotemark{b}} &
\colhead{Mar 2004\tablenotemark{c}} &  
\colhead{Dec 2004\tablenotemark{c}} \\
\colhead{(Fig. 5)} &
\colhead{Number} &
\colhead{WFPC2} &
\colhead{WFPC2} &
\colhead{ACS} &
\colhead{ACS} }
\startdata
A & 1228 & 7.3 $\pm 1.1$    & 9.6 $\pm 0.7$  & 8.2 $\pm 0.6$   & 3.5 $\pm 0.4$ \\
B & 1273 & $<$1.5 ~~~~      & 4.4 $\pm 0.6$   & 4.9 $\pm 0.5$   & 2.2 $\pm 0.3$ \\
C & 1296 & 5.2 $\pm 0.7$    & 23.9 $\pm 1.3$  & 13.1$\pm 0.6$   & 5.0 $\pm 0.5$ \\
D & 1297 & 3.2 $\pm 0.6$    & 2.4 $\pm 0.7$   & 2.8 $\pm 0.4$   & 7.6 $\pm 0.6$ \\
E & 1323 & 2.7 $\pm 0.5$    & 1.5 $\pm 0.4$   & 5.3 $\pm 0.3$   & 4.9 $\pm 0.4$ \\
F & 1325 & 4.7 $\pm 0.5$    & 7.8 $\pm 0.5$   & 2.4 $\pm 0.2$   & 2.0 $\pm 0.3$ \\
G & 1330 & 5.5 $\pm 0.8$    & 12.1 $\pm 0.6$  & 6.7 $\pm 0.4$   & 5.1 $\pm 0.4$ \\
H & 1344 & 17.4$\pm 2.1$    & 2.7 $\pm 0.5$   & 4.0 $\pm 0.2$   & 3.4 $\pm 0.4$ \\
\enddata
\tablenotetext{ ~ }{Fluxes are in units of 10$^{-17}$ erg cm$^{-2}$ s$^{-1}$. }
\tablenotetext{a}{From the Cas A outer ejecta knot catalog of \citet{HF08}.}
\tablenotetext{b}{Measured from WFPC2 F675W images.}
\tablenotetext{c}{Measured from combined ACS F625W + F775W images.}
\end{deluxetable*}

Standard pipeline data reduction of these images was performed using
IRAF/STSDAS\footnote{IRAF is distributed by the National Optical Astronomy
Observatories, which is operated by the Association of Universities for
Research in Astronomy, Inc.\ (AURA) under cooperative agreement with the
National Science Foundation. The Space Telescope Science Data Analysis System
(STSDAS) is distributed by the Space Telescope Science Institute.}.  This
included debiasing, flat-fielding, geometric distortion corrections,
photometric calibrations, and cosmic ray and hot pixel removal.  The STSDAS
{\it {drizzle}} task was also used to combine single exposures in each filter.

For the WFPC2 F675W images, fluxes for eastern limb knots were measured using
the IRAF tasks `apphot'.  For the ACS F625F, F775W, and F850LP images,
calibrated estimates for fluxes of each outer optical ejecta knot for the three
filters were taken from the outer ejecta knot catalog of \citep{HF08} which
were calculated using the SExtractor \citep{Bertin96} automated source
extraction software package.  For cases where the automated programs failed,
knot fluxes were calculated manually five pixel wide apertures. When fluxes were
manually calculated, background estimation was performed by calculating the
total $5$ pixel aperture flux in at least five positions near the object
(avoiding other sources) and then subtracting the mean computed ``background''
flux from the total object pixel sum. Final detector count numbers were then
converted to flux by multiplying by the mean flux density per unit wavelength
generating 1 count s$^{-1}$ (i.e., the `PHOTFLAM' factor) times the filter
effective bandwidth (EBW).

Due to the non-spherical nature along with sometimes changing morphology of
some ejecta knots, plus the closeness of adjacent emission knots, flux
measurements for many of these outlying ejecta knots are subject to greater
errors than simply photometric accuracy. Moreover, differences between filter
bandpass and system throughputs between the WFPC2 F675W filter and the combined
ACS F625W + F775W adds to the uncertainty when comparing January 2000 and
January 2002 WFPC2 fluxes to March and December 2004 ACS fluxes.  Nonetheless,
since the WFPC2 F675W filter images and the ACS F625W + F775W  filter images
cover the same ejecta emission lines, namely [\ion{O}{1}] $\lambda\lambda$6300,
6364, H$\alpha$, [\ion{N}{2}] $\lambda\lambda$6548, 6583, [\ion{S}{2}]
$\lambda\lambda$6716, 6731, and [\ion{O}{2}] $\lambda\lambda$7319, 7330 (see
\citealt{Fesen01} for further details), we have attempted to investigate some
knot flux changes over the nearly 5.5 yr interval (June 1999 to December 2004)
using these two {\sl HST} image sets. 

For our investigation of relative flux differences between different epochs using 
both WFPC2 and ACS images, detector counts were summed using $5 \times 5$ to
$9 \times 9$ pixel sized windows depending on knot size and neighboring
emission complexity, subtraction of a local mean background, and finally
normalized to the measured flux of several neighboring field stars.  Flux
errors listed in Tables 1 and 2 include apphot measurement uncertainties,
background non-uniformities, and measurement deviations observed when varying
knot centroids.  While most 1 $\sigma$ uncertainties listed are below 10\%,
measured flux uncertainties can be as much as 30\% for some knots.

\begin{figure*}
\centering
\includegraphics[width=0.7\linewidth]{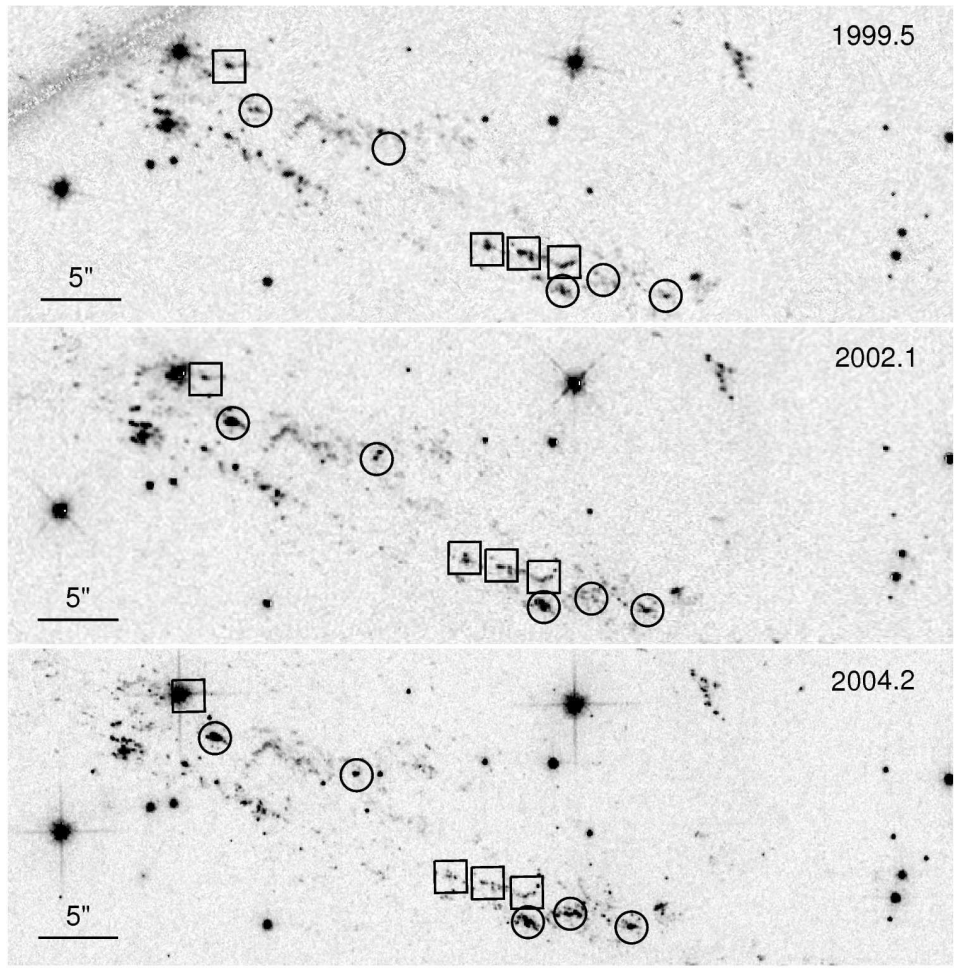}
\caption{{\sl HST} WFPC2 F675W images (top two panels) and a combined ACS F625W + F775W image (bottom panel)
of one of the streams of ejecta knots in Cas A's northeastern `jet' of ejecta. Images cover the
time period of 1999.5 to 2004.2 with a time interval between images of roughly two years. 
A significant fraction  of ejecta knots exhibit changes in brightness, with
circles and boxes marking knots showing flux increases or decreases, respectively. 
The approximate center of field-of-view
of the panels is: $\alpha$[J2000] = $23^{\rm h} 23^{\rm m} 49.4^{\rm s}$,
$\delta$[J2000] = $58^{\rm o} 49' 50''$).
North is up, East to the left.  }
\label{fig:jet_stream}
\end{figure*}

\section{Results}

In general, we found that a significant fraction of ejecta knots around all
portions of the outer periphery of Cas A show some degree of optical flux
variability.  Below, we discuss these flux changes along with morphological
changes in some of these outer ejecta knots. Due to differences in expansion
velocities and possible chemical compositions, the following discussion is
divided into sections by knot location.   

\subsection{Northeast Jet Knot Variability}

Figure 1 shows {\sl HST} images for one of the streams of ejecta in
the remnant's NE jet feature taken at three epochs covering a time span of
nearly five years.  The top panel is a June 1999 WFPC2 F675W image, the middle
panel a January 2002 WFPC F675W image, with the bottom panel a March 2004 F625W
+ F775W ACS image.  All three images are sensitive essentially to the same set
of emission lines; namely, [\ion{O}{1}] $\lambda\lambda$6300, 6364,
[\ion{O}{2}] $\lambda\lambda$7319, 7330, and [\ion{S}{2}] $\lambda\lambda$6716,
6731. Thus changes seen between these images should be representative of
overall flux changes in a knot's line emission. 

As Figure 1 illustrates, large brightness changes occur in many of the NE jet's
ejecta knots across the whole length of this ejecta stream during this five
year period.  Ejecta knots and knot complexes exhibiting especially large flux
increases or decreases have been marked in the figure by circles and boxes,
respectively.  Despite significant changes in many individual knots and knot
clusters, the overall optical appearance of this stream of ejecta does not
dramatically change over this nearly five year time period.

Significant flux changes for some NE jet knots can occur over much shorter
timescales and we have investigated whether such changes are associated with
varying elemental abundances or changes in the ionization state of the ejecta.
To do this, we examined several NE jet ejecta knots using ACS images taken nine
months apart (March and December 2004) using the F625W, F775W and F850LP
filters.  These filters are primarily sensitive to [\ion{S}{2}]
$\lambda\lambda$6716, 6731, [\ion{O}{2}] $\lambda\lambda$7319, 7330 and
[\ion{S}{3}] $\lambda\lambda$9069, 9531 line emission, respectively. They thus
serve as flux variability indicators sensitive to both knot chemistry and
ionization state.

\begin{figure*}[hpt!]
\centering
\includegraphics[width=\linewidth]{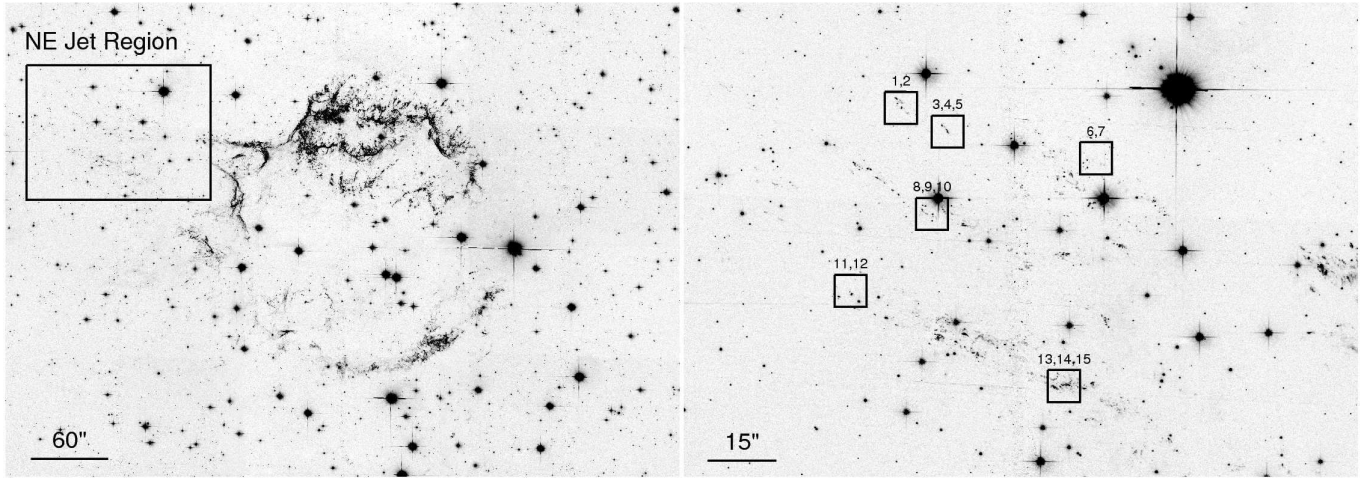}
\caption{December 2004 {\sl HST} ACS F625W + F775W image of the Cassiopeia A supernova remnant. 
Left panel marks the area of the remnant's northeastern jet of ejecta knots which is enlarged
(right panel) with the six selected areas shown in Figure 3.  }
\label{fig:jetknotmap}
\end{figure*}

\begin{figure*}[hp!]
\centering
\includegraphics[width=0.46\linewidth]{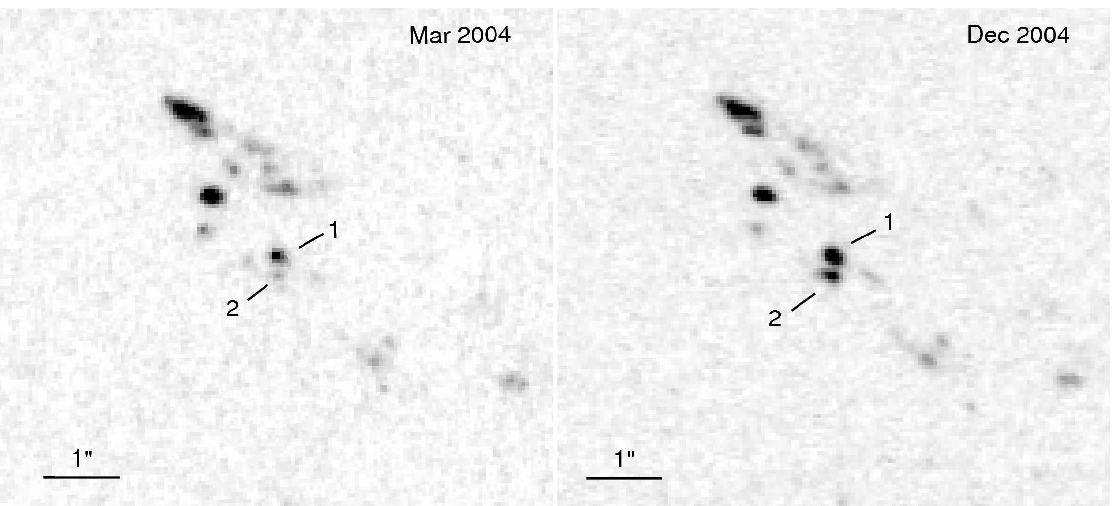}
\includegraphics[width=0.46\linewidth]{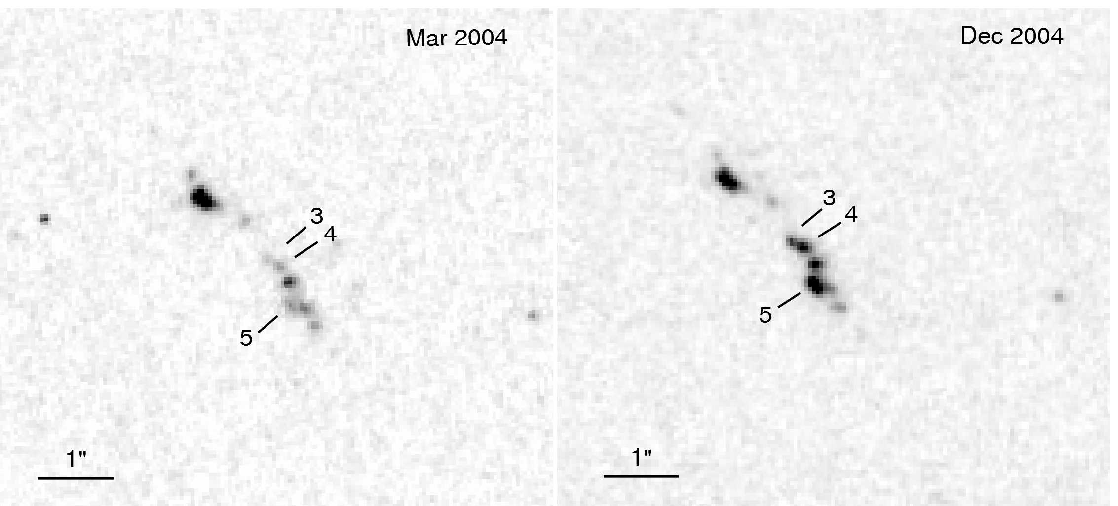}\\
\includegraphics[width=0.46\linewidth]{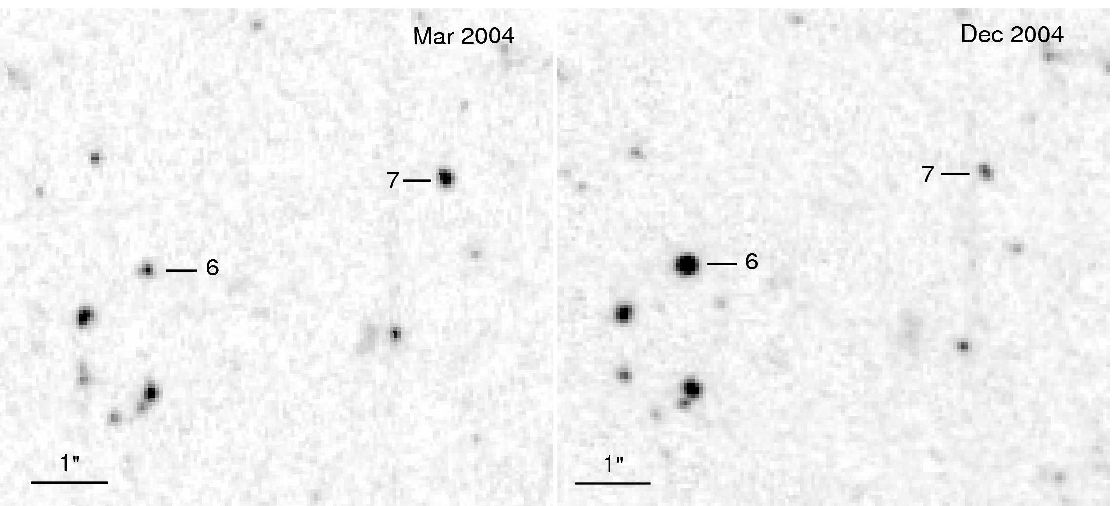}
\includegraphics[width=0.46\linewidth]{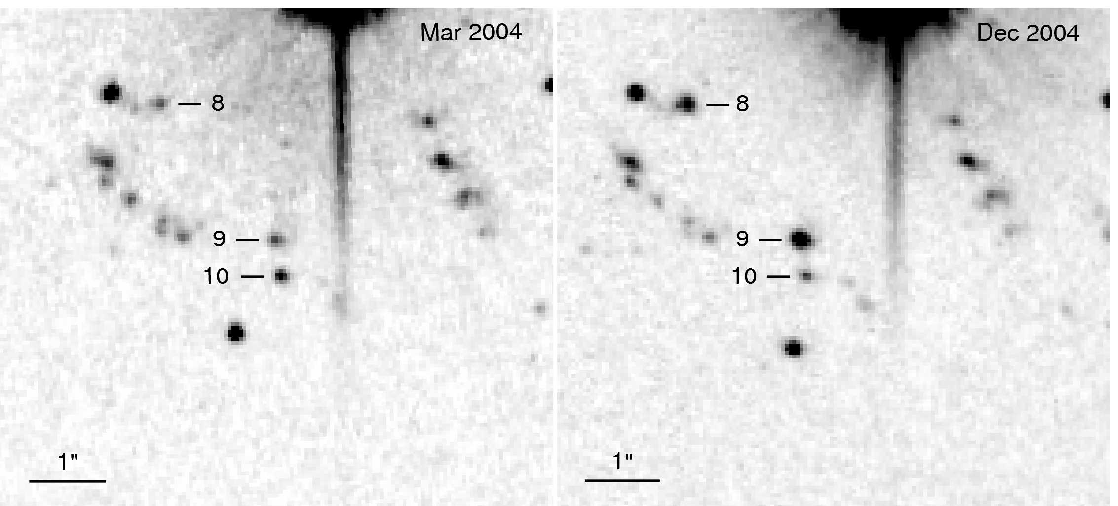}\\
\includegraphics[width=0.46\linewidth]{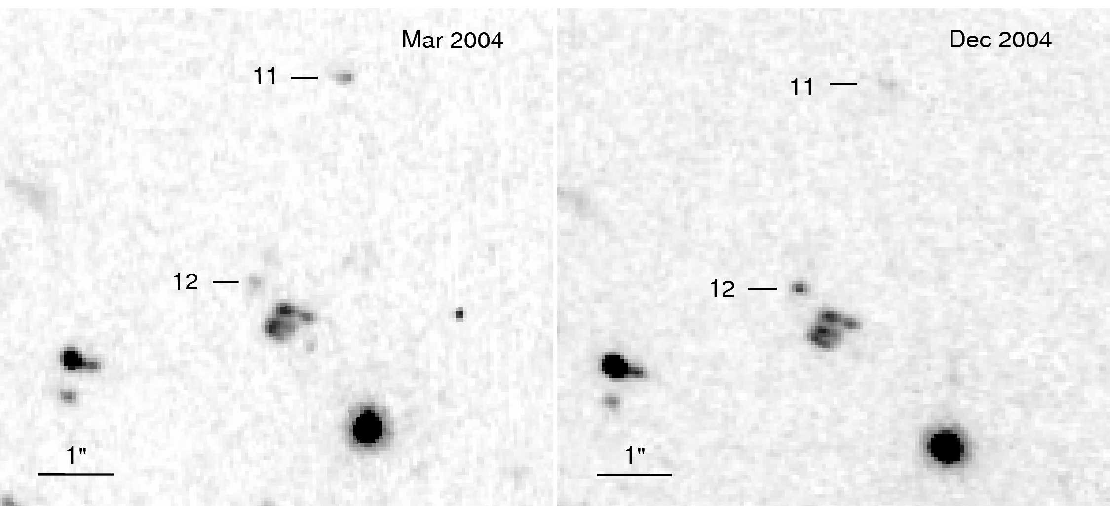}
\includegraphics[width=0.46\linewidth]{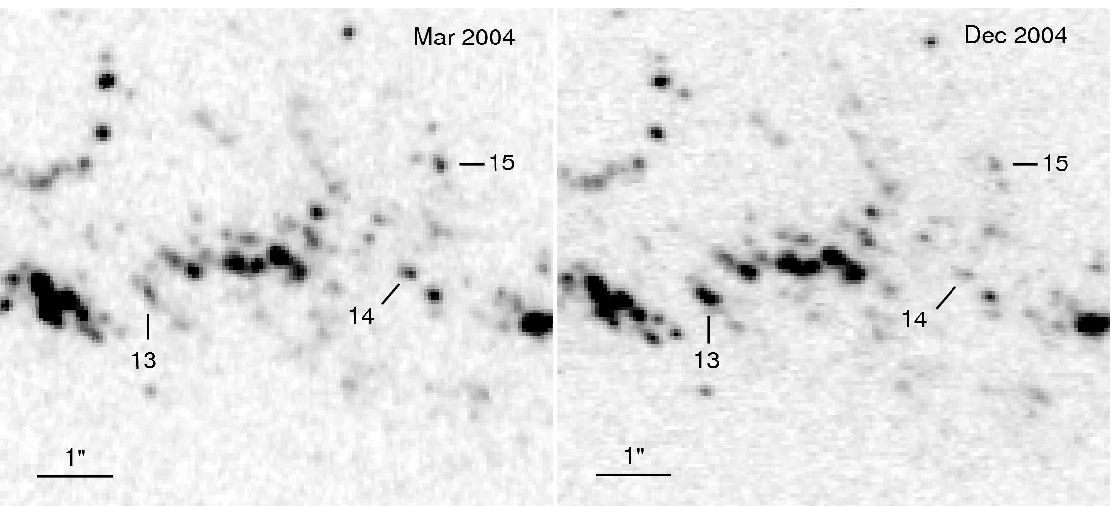}
\caption{March and December 2004 ACS F625W + F775W images
showing flux changes in ejecta knots over a nine month timespan.  }
\label{fig:jetknots}
\end{figure*}

The location of 15 selected knots in the remnant's NE jet which were found to
exhibit large brightness changes over this nine month period are shown in
Figure 2, both in relation to the whole remnant and within the NE jet region.
Enlargements of the March and December 2004 ACS F625W + F775W images of these
15 knots are shown in Figure 3. To study possible abundance and ionization
level changes, we measured F775W (i.e., [\ion{O}{2}]) and F850LP (i.e.,
[\ion{S}{3}]) fluxes for March and December 2004 for these knots are listed in
Table 1.

Knots that brightened optically over this period are $1 - 6$ and 8, 9,
12, and 13.  Conversely, Knots 7, 10, 11, 14 and 15 decreased in
flux. Overall, a majority of knot flux changes involved brightening,
with only a small percentage of knots showing flux decreases over the March to
December 2004 time period.  While our investigation concentrated on significant
knot flux changes, knots that brightened were found to increase in flux by
larger factors ($\simeq3 - 8$) as compared to that seen for knots that faded.
Although the five knots listed in Table 1 which faded between March and
December 2004 appear to fade by nearly the same factor of $\sim$2 suggesting
similar decay light curves, examination of other jet knots indicate a wider
range of flux decrease is possible. For example, the \citet{HF08} catalog knots
419, 995, and 1019 showed 2004 Dec/Mar F775W flux ratios of 0.35, 0.25 and
0.35, respectively.

In general, we detected no pattern in terms of brightness changes between the
three filters to suggest systematic changes in the line emissions
between knot brightening or fading.  This can be seen in Table 1 where we
compare the measured F775W and F850LP fluxes which are primarily sensitive to
[\ion{O}{2}] and [\ion{S}{3}], respectively. (Note: We excluded flux
measurements in the F625W filter in Table 1 since it is sensitive to both
[\ion{O}{1}] $\lambda\lambda$6300, 6364 and [\ion{S}{2}] $\lambda\lambda$6716,
6731, thus complicating interpretation of any observed flux changes.)

While large brightness changes were only seen in a minority of jet emission
knots ($<$25\%), smaller flux changes were apparent in a much larger percentage
of knots, approaching 50\% when viewed over a longer time frame (see Fig.\ 1).
However, with few exceptions, faint emission was nearly always visible prior to
or following even large brighten or fading episodes suggesting a minimum
excitation level of the ejecta knot outside of these large flux change
occurrences.  

\begin{figure*}
\includegraphics[width=\linewidth]{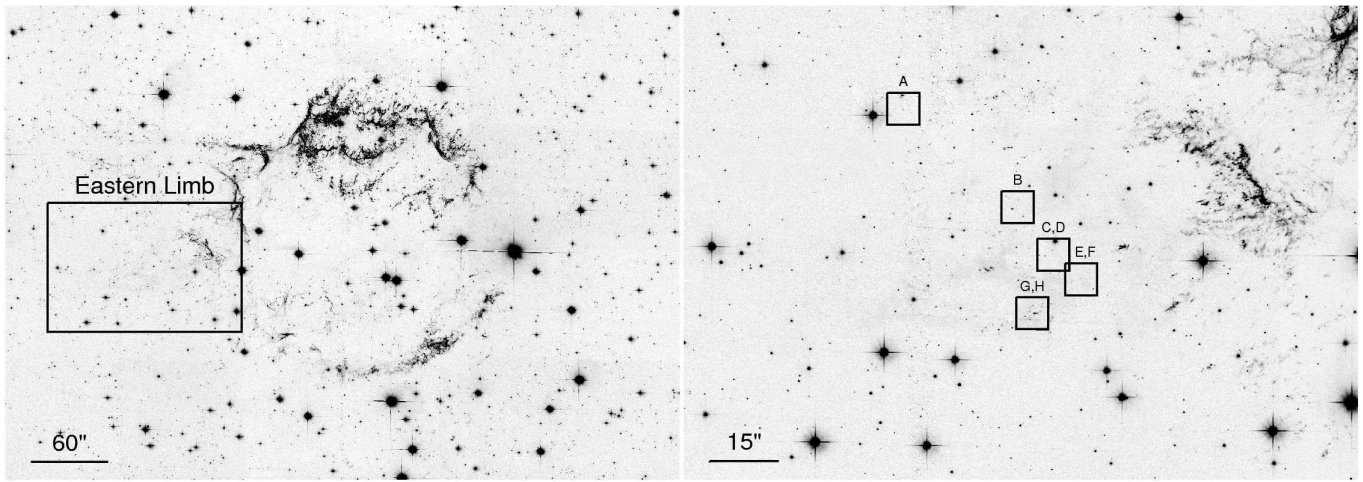}
\caption{December 2004 {\sl HST} ACS F625W + F775W
images of the eastern limb of the Cassiopeia A supernova remnant. Left panel shows the region examined for
outer ejecta knot flux variability indicated by the box. The right panel shows the small regions
enlarged in Figure 5. }
\label{fig:eastknotmap}
\end{figure*}

\subsection{Knot Variability along the Eastern Limb}

Substantial knot brightness changes were also seen in the remnant's
high-velocity ejecta located along its eastern limb.  Unlike the NE jet
knots, most ejecta here exhibit a spectrum dominated by H$\alpha$ and
[\ion{N}{2}] emission lines, indicating a less O and S rich composition
\citep{Fesen01,HF08}.

The portion of the eastern limb region that was examined for flux changes
is shown in the left panel of Figure 4.  Locations of selected ejecta knots showing
large brightness changes are marked in the five smaller regions indicated in
the right panel of Figure 4, with 2000 and 2002 WFPC2 and 2004 ACS enlargements of
these regions presented in Figure 5.  Within these smaller regions, flux
measurements were made of eight individual knots (knots A - H; see Table 2). 

A variety of light curves was found here, with some knots going up and down in
brightness (`flickering') in times as short as one year. For example, Knot A
brighten slightly between January 2000 and January 2002 but then sharply
declined between March and December 2004 (see Fig.\  5, top row of images).
Knot B (second row from top) was undetected in the January 2000 WFPC2 image
becoming visible in January 2002 but then faded again by December 2004.
Whereas Knot G (bottom row) underwent similar flux changes as Knot B,  Knot C
(middle row) was more extreme, increasing four-fold between 2000 and 2002 but
then fading by a similar factor by December 2004.

In contrast, some ejecta knots such as Knot D were nearly constant in flux
(within measurement uncertainties) over nearly four years (Jan 2000 -- Mar
2004) but then showed a substantial brightening over just nine months (March to
December 2004).  Similarly, Knot F (second row from bottom) showed an abrupt
decrease in brightness between 2002 and 2004.  

Overall, an array of seemingly random brightness variability, both large and
small, was observed not unlike that observed in the NE jet ejecta, except that
here we actually observe flickering.  While not all outlying
eastern limb ejecta knot exhibit large brightness variations over the nearly
five year time frame covered by these {\sl HST} images, many knots did.
Increases or decreases in brightness by 25\% or more were not uncommon.

\begin{figure*}
\centering
\includegraphics[width=0.7\linewidth]{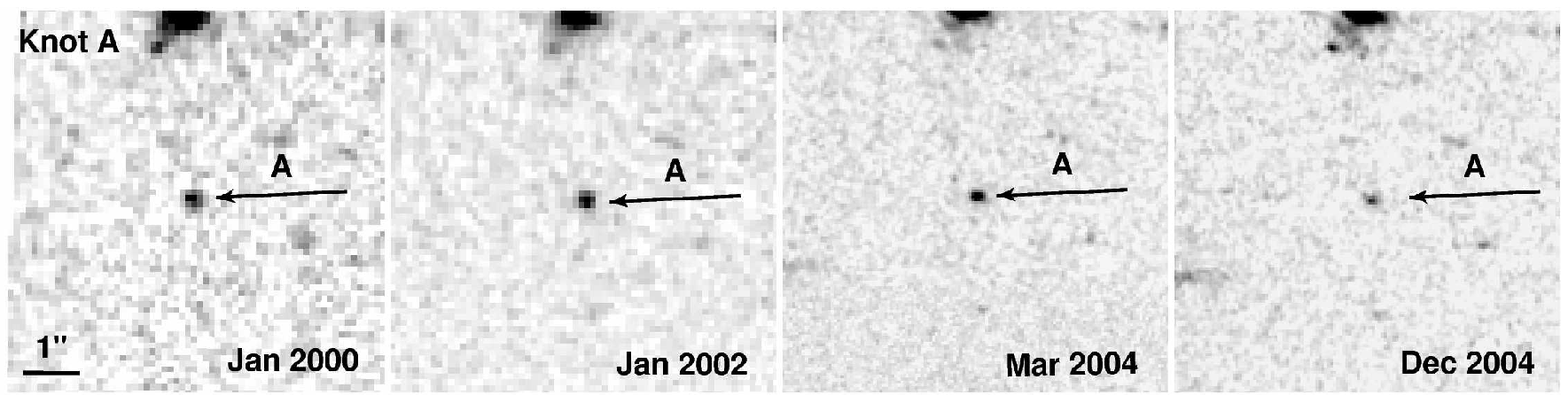}\\
\includegraphics[width=0.7\linewidth]{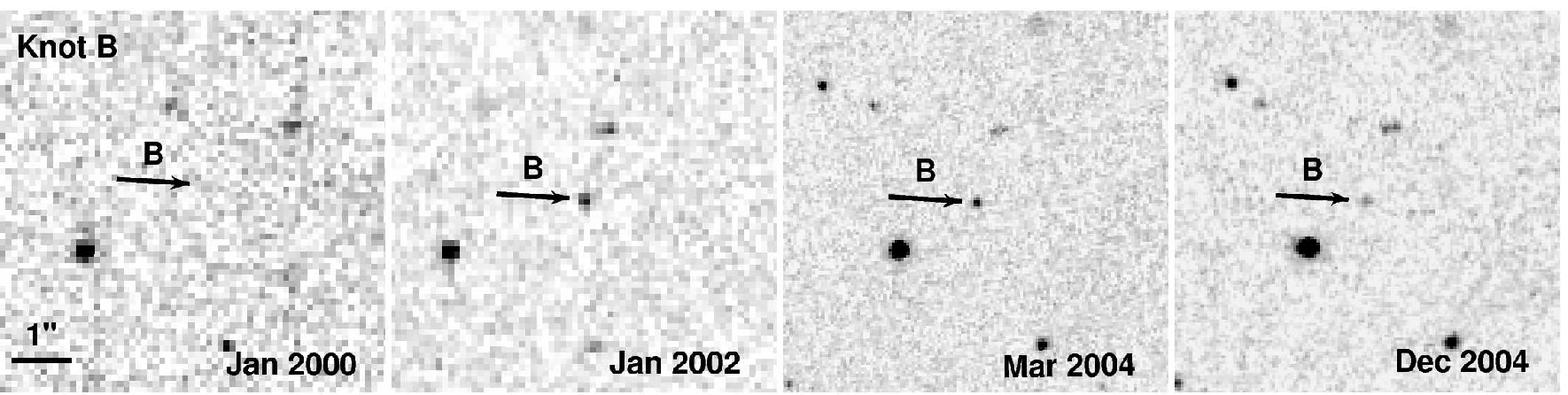}\\
\includegraphics[width=0.7\linewidth]{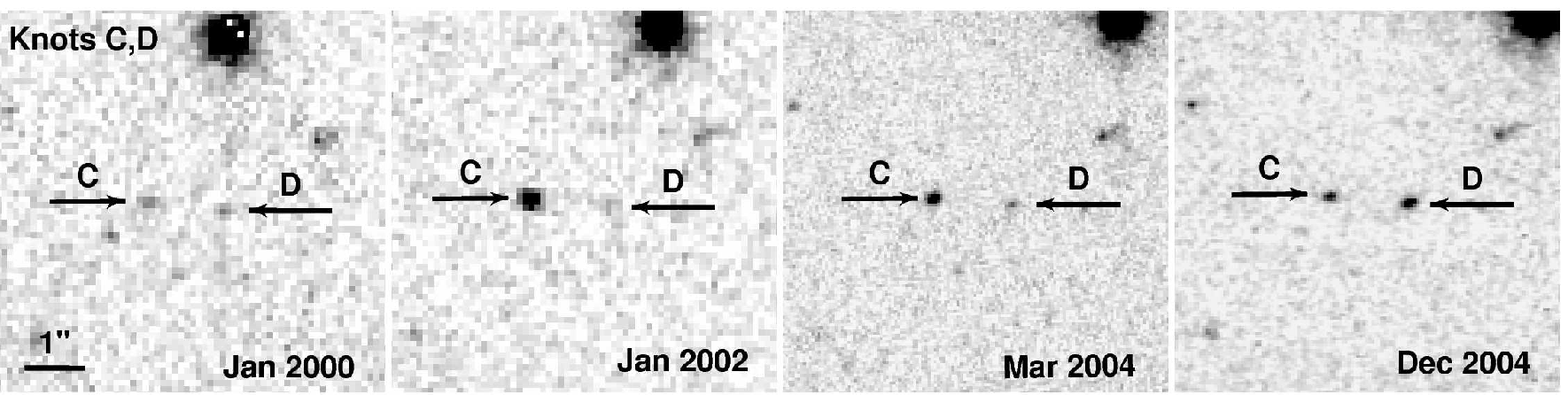}\\
\includegraphics[width=0.7\linewidth]{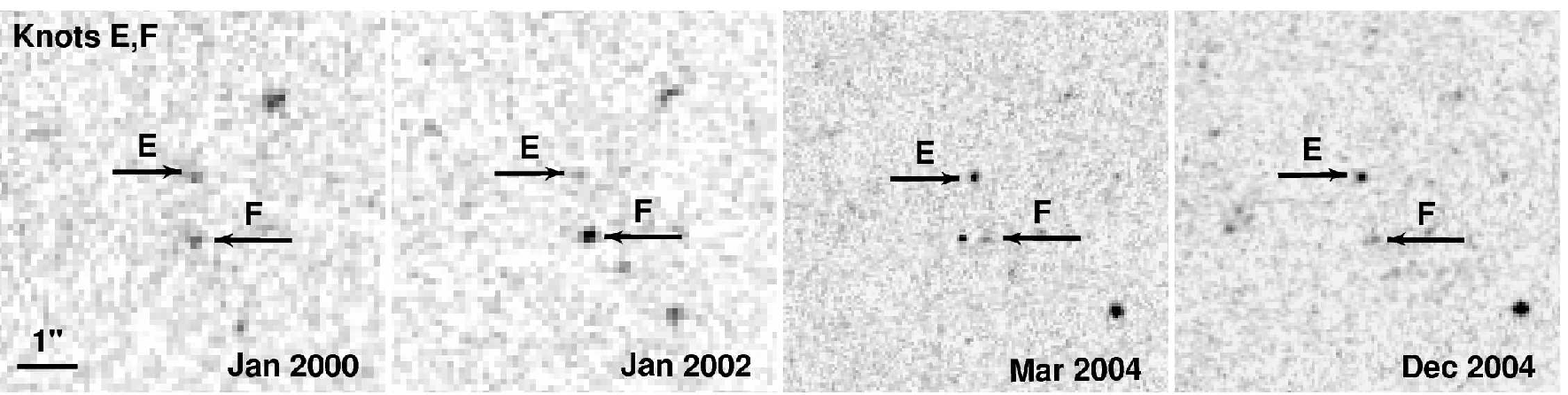}\\
\includegraphics[width=0.7\linewidth]{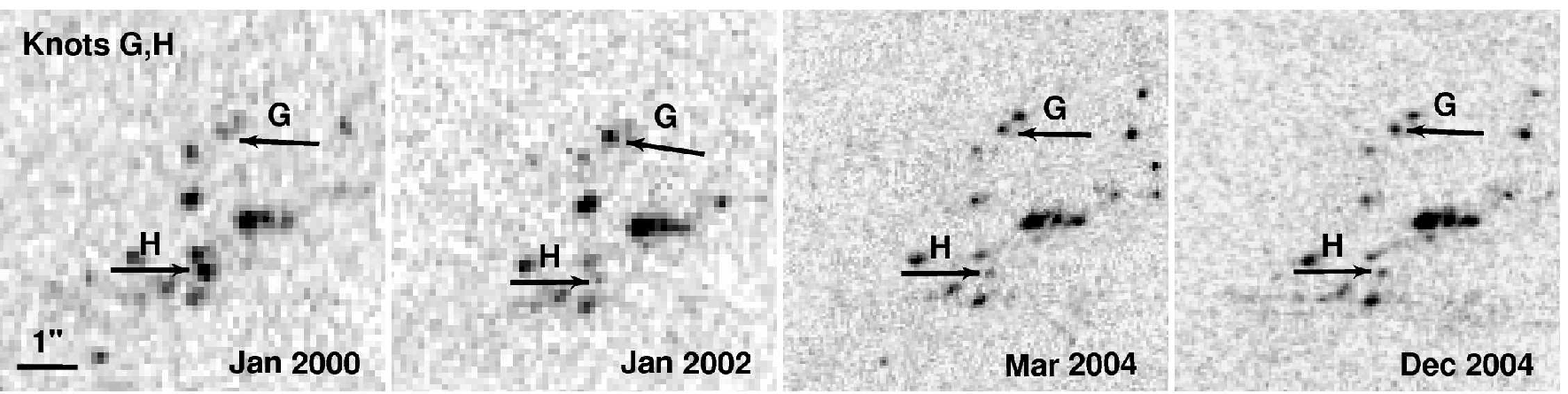}
\caption{Enlarged 2000 and 2002 WFPC2 images and 2004 ACS images of five regions
along Cas A's eastern limb highlighting several SN ejecta knots showing significant
flux variations. Measured filter fluxes for knots marked A through G are listed in
Table 1.   }
\label{fig:knota-g}
\end{figure*}

Although most brightness changes appear uncoordinated, a few cases do suggest
sequential brightening such as the closely spaced knots C and D (Fig.\ 5,
middle row).  These two knots are located nearly directly east of the Cas A
expansion center with corresponding motions nearly due eastward.  As noted
above, Knot C significantly brightened between 2000 and 2002 but then sharply
faded by March 2004.  Knot D, following closely behind Knot C as seen in
projection (separation = $1\farcs3$), showed little change in flux between 2000
and March 2004 but a sharp increase in brightness by December 2004.  

As illustrated by the vertical line shown in Figure 6, the location in the sky
were both knots C and D brightened is nearly the same within $0\farcs5$
suggesting they may have sequentially encountered a circumstellar or
interstellar cloud (not visible on these {\sl HST} images) at this spot which
led to their shock excitation and hence optical brightening. This is consistent
with their observed $1\farcs3 \pm 0\farcs1$ separation and their measured
proper motion of $0\farcs434$ per year \citep{HF08}.  These values imply that
Knot D would have reached Knot C's projected January 2002 position some 3.0
years later and this is exactly what is observed.  Knot C brightened in January
2002 roughly 3 years before Knot D was observed to brighten in December 2004.

\begin{figure}
\includegraphics[width=\linewidth]{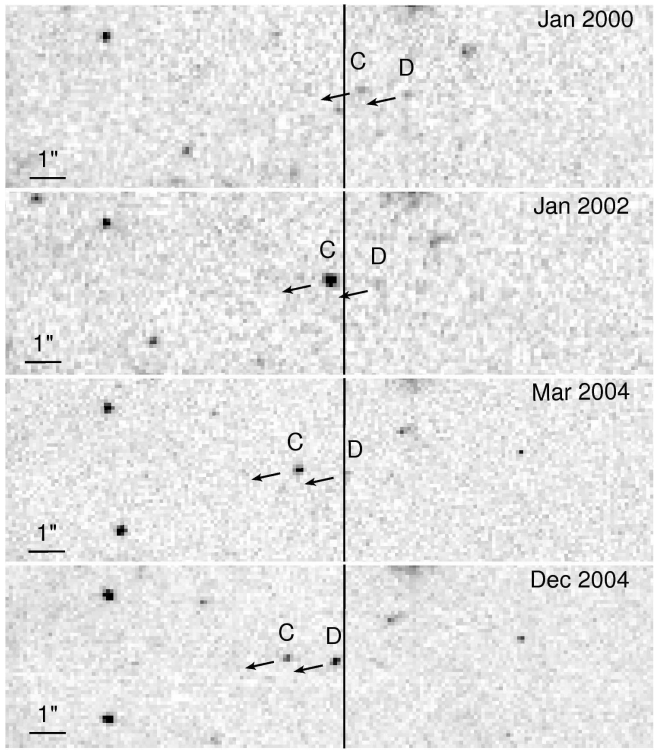}
\caption{Flux variations of high-velocity eastern limb knots C and D as they move in the directions
indicated by the arrows. The vertical line indicates the location
 where each knot abruptly brightens, suggestive of an encounter with a circumstellar or interstellar
cloud which is invisible in these images. North is up, East to the left. }
\label{fig:Knots_CnD.ps}
\end{figure}

\subsection{Knot Emission Tails}

The {\sl HST} images of Cas A's outer ejecta knots also revealed a small
percentage of knots ($\lesssim$ 5\%) with extended, trailing emission
structures or `tails' typically $0\farcs2 - 0\farcs7$ in length.  Such trailing
emission features were found in knots all across the remnant's outer
periphery.  Figure 7 shows the locations of ten representative examples of
knots with trailing emissions, with enlargements of combined December 2004 ACS
F675W + F775W images for these knots presented in Figure 8.

\begin{figure}
\centering
\includegraphics[width=\linewidth]{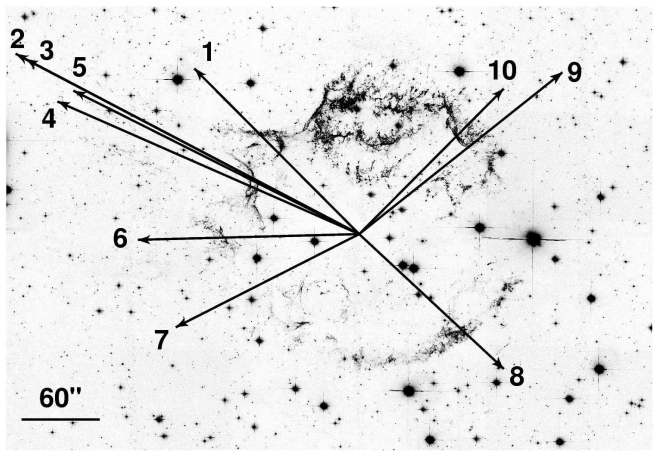}
\caption{Combined December 2004 {\sl HST} ACS F625W + F775W image of the Cassiopeia supernova remnant.
The arrows originate from the remnant's estimated expansion center \citep{Thor01} with their tips marking the 
locations of ten examples of outlying, high-velocity ejecta knots which appear to exhibit faint trailing emission `tails'
as shown in the enlargements presented in Figure 8. }
\label{fig:tailmap}
\end{figure}

\begin{figure}
\includegraphics[width=\linewidth]{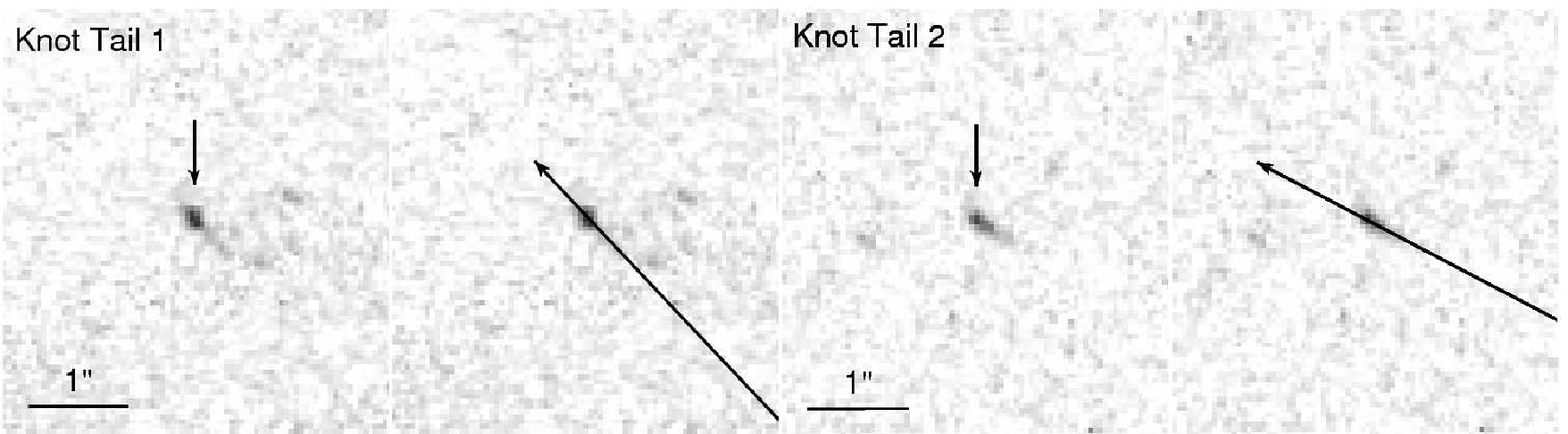}
\includegraphics[width=\linewidth]{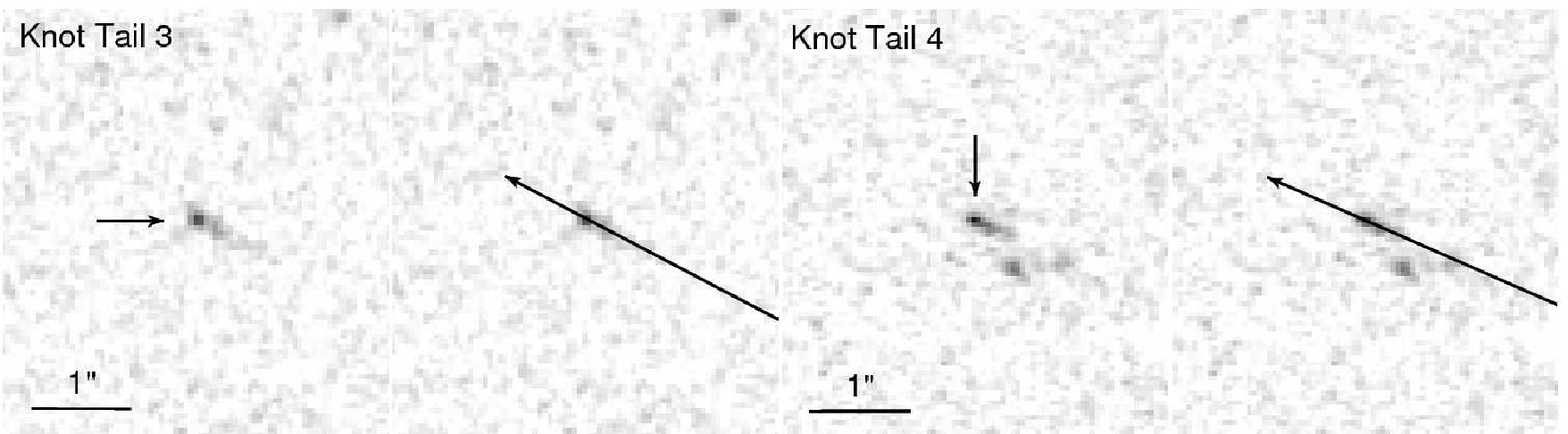}
\includegraphics[width=\linewidth]{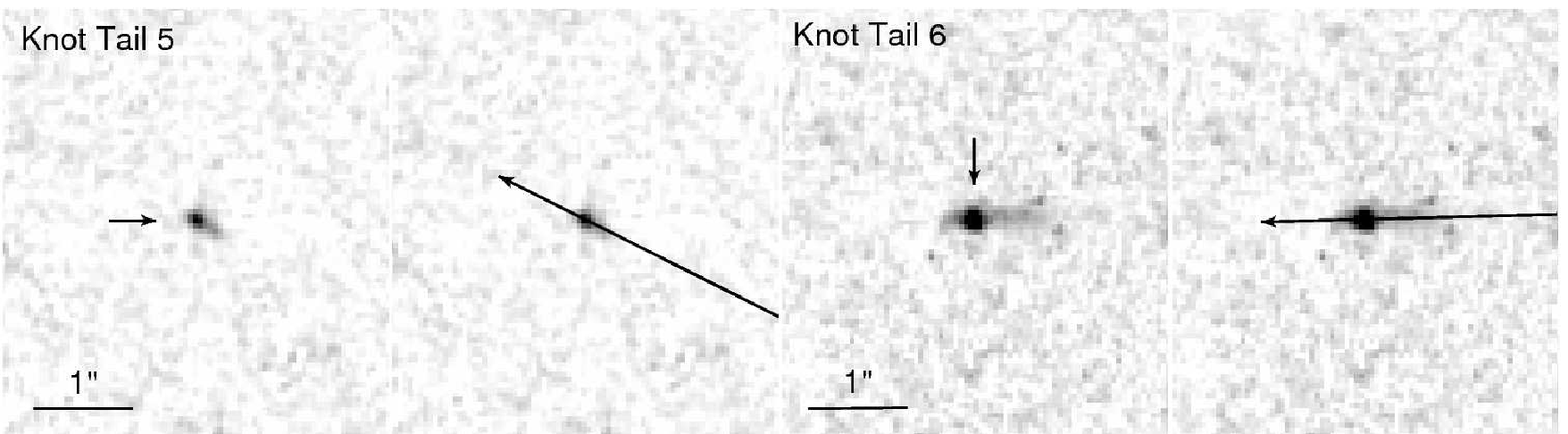}
\includegraphics[width=\linewidth]{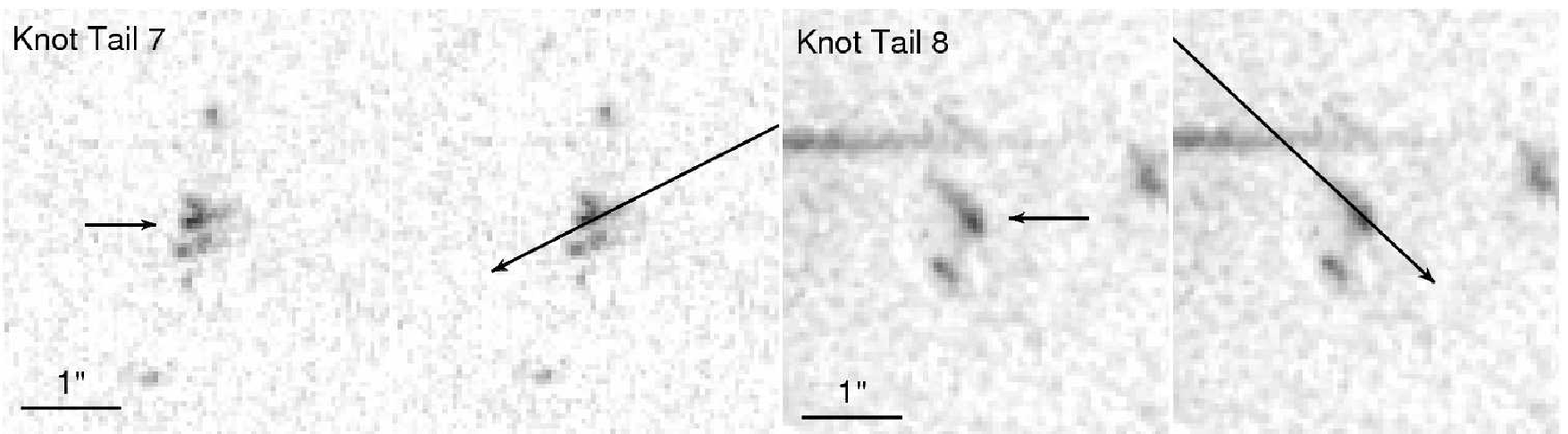}
\includegraphics[width=\linewidth]{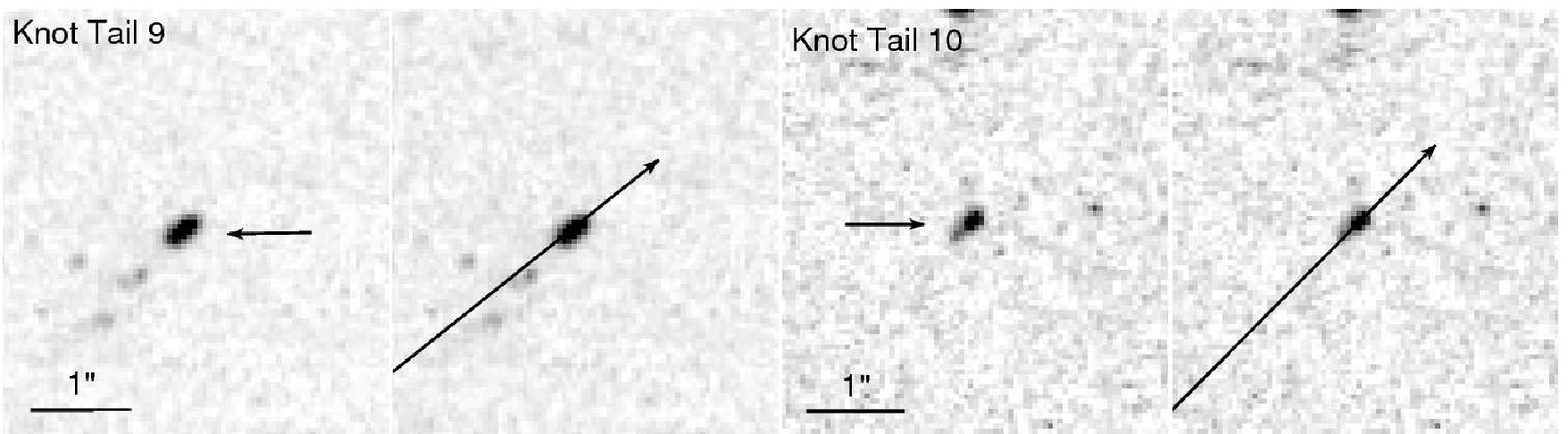}
\caption{Enlargements of the combined December 2004 ACS F675W + F775W image
showing extended emission behind several outlying ejecta knots.
Each knot is shown with and without an arrow marking its direction of motion from the
remnant's estimated center of expansion.  }
\label{fig:knot_tails}
\end{figure}

Trailing emissions were always found to be in excellent alignment with the
knot's estimated direction of motion strongly suggestive of mass stripping.
This tail -- knot motion alignment is illustrated in Figure 8 where each knot
is shown twice, with (right panel) and without (left panel) an arrow drawn from
the estimated remnant's center of expansion \citep{Thor01} out to the knot's
December 2004 location.  In some cases, the trailing emission appears smooth in
intensity away from the knot's head (e.g., Knots 1, 5, 6 and 7), whereas in
other cases the trailing emission appears to show one or more trailing emission
knots (e.g., Knots 2, 3, and 9). 

In principle, these emission tails could be in part or wholly due to
shock-excited CSM or ISM emission caused by the passage of these high-velocity
ejecta. But this seems unlikely. Comparisons of the knot fluxes in the ACS
filters F625W, F775W, and F850LP, shows no significance difference between a
knot's head or its tail in terms of relative strengths. For example, if the tails were due
to excited CSM or ISM, H$\alpha$ and [\ion{N}{2}] $\lambda\lambda$6548,6583
would then be expected to be the main emission contributors making the extended
trailing emission mainly visible in the F625W images. However, this is not what
is observed; trailing emissions are visible in all three filter images
consistent with the picture of mass ablation and/or partial knot disruption. 

\begin{figure}
\includegraphics[width=\linewidth]{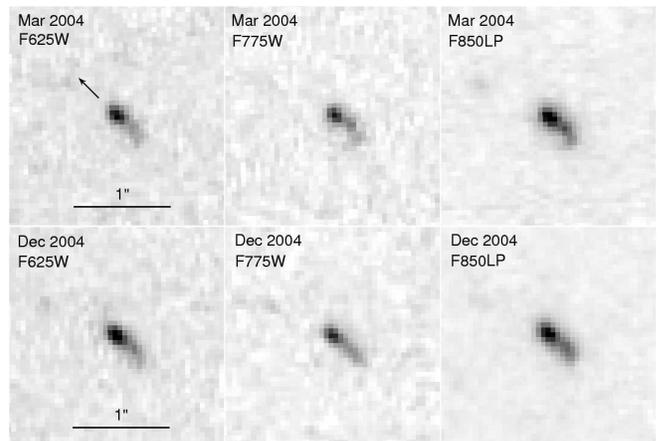}
\caption{March and December 2004 {\sl HST} ACS F625W, F775W, and F850LP filters images 
         of the ejecta knot No.\ 429 (\citealt{HF08}) in the NE Jet showing 
         a prominent head-tail structure. The arrow in the upper left frame 
         indicates the knot's direction of motion. North is up, East is to the left.}
\label{fig:knot_429_tail_images}
\end{figure}

A closer examination of a knot's trailing emission is shown in Figure 9 where
we present {\sl HST} image enlargements of a NE jet ejecta knot with a bright
emission tail $\simeq 0\farcs5$ long.  This knot (No.\ 429; \citealt{HF08})
lies in the central region of the NE jet. Intensity plots shown in Figure 10
indicate the trailing emission is some 50\% -- 75\% as bright as the knot head
itself. In order to generate this level of emission, the material in the tail
likely represents a substantial amount of mass. 

The length and structure of the knot's trailing emission also changes slightly
between the March and December images. For example, the knot's emission tail
shows the presence of an emission clump in the trailing emission.  The clump's
position relative to the knot's head changes between March and December in both
the F625W and F850LP images amounting to a shift of $\simeq0\farcs05$ opposite
of the knot's outward direction of motion.

The knot has a measured proper motion of $0\farcs60$ yr$^{-1}$ \citep{HF08}
which converts to a transverse velocity of $\simeq$ 9700 km s$^{-1}$ at a
distance of 3.4 kpc. Thus, the smaller forward proper motion of the tail knot
implies a lower transverse velocity by $\simeq$ 800 km s$^{-1}$. This velocity
difference, together with it relative brightness, suggests a deceleration of
some portion of the main ejecta knot over this nine month time frame.

\begin{figure}
\includegraphics[width=\linewidth]{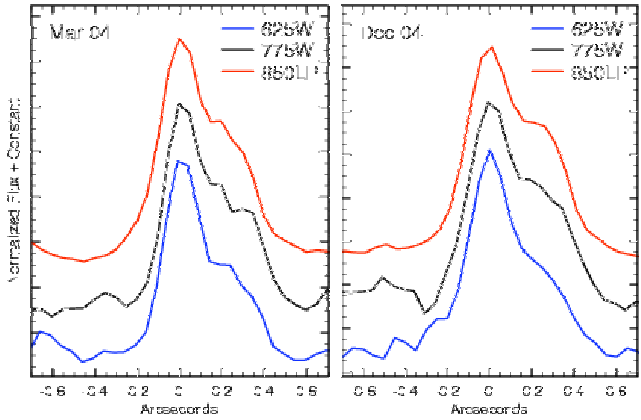}
\caption{Observed head-to-tail flux plots for the NE Jet ejecta knot No.\ 429 shown in Fig.\ 9.} 
\label{fig:knot_429_tail_plots}
\end{figure}

\section{Discussion}

Ejecta knot flickering, ablation tails, and fragmentation are expected
phenomena associated with the gradual dissolution of ejecta knots caused by
their high-velocity interaction with an inhomogeneous circumstellar or
interstellar medium.  In the case of Cas A's outer ejecta whose chemical
abundances are significantly non-solar, such changes are subject to very
different time scales and scale lengths than seen in other astrophysical
settings; e.g., HH objects in T-Tauri outflows.  

Several groups have analyzed the dynamical interaction between SN ejecta and a
surrounding low density medium
\citep{Chevalier75,Hamilton85,Jones94,Anderson94,Cid96,B02,WC02}.  A standard
feature of this interaction is that a slower shock is driven into the ejecta
knot, which subsequently undergoes compression and lateral deformation.  

The knot's internal shock heating and subsequent cooling via radiative
processes can generate substantial optical line emission.  For Cas A's outlying
ejecta, the plasma's cooling rate will be enhanced significantly by their 
metal-rich abundances with additional cooling possible via metal ions sputtered off
dust grains (for discussions of dust in Cas A ejecta see
\citealt{Lagage96,Arendt99,Rho08}).

The supersonic motion of the knot through the ambient medium will form a bow shock
upstream of the cloud and a strong shear layer will develop at the knot's surface
which is subject to the Kelvin-Helmholtz instability \citep{MacLow94}.  The
cloud will become compressed by the internal shock with its boundaries
distorted by Rayleigh-Taylor and Richtmyer-Meshkov instabilities and by vortex
generation as hot gas flows around the knot's edge.  

In model simulations of a shocked ISM cloud, a cloud is typically
destroyed in several dynamical ``cloud-crushing times'', $t_{\rm cc}$, defined as
the transit time for the internal shock to move entirely through the knot 
\citep{Klein94,Jones94,Pol04b}.  The end result is that the cloud becomes
fragmented, decelerated, and eventually shredded.

One can gain some understanding of the relevant time scales for the interaction
and microphysical processes with the following estimates.  We consider a SNR
ejecta knot moving at velocity $V_{k}$ with respect to the ambient SNR gas,
of density $n_{\rm e} \approx1$ cm$^{-3}$ \citep{Braun87,CO03,HL09}.  The
knot-to-ambient density contrast ratio, $\chi$, typically ranges from
$10^2$--$10^4$ \citep{Fesen01}.  The internal shock moves through the knot at
velocity $V_s \approx \chi^{-1/2} \, V_{k}$.  Assuming $V_{k} = 5000 -
10,000$ \kms\ and $\chi \approx 100-10,000$, models show the internal shocks
propagate at $V_s \approx 100 - 500$ \kms  \citep{Silvia10}.  

The strong adiabatic shock compresses the gas, initially by a factor of 4, and
heats the post-shock gas to temperatures between $10^5$~K and $10^7$~K.  The
Rankine-Hugoniot temperature jump at the shock front is given by:
\begin{equation} T_s =  \frac {3}{16} \left( \frac {\mu \, V_s^2}  {k} \right)
= (1.4 \times 10^5~{\rm K}) \left( \frac {\mu}{0.6 m_H} \right) \, V_{100}^2
\end{equation} for shocks of velocity $V_s = (100~{\rm km~s}^{-1}) V_{100}$ and
ionized gas with mean molecular weight $\mu \approx ~ 0.6m_H$.

One can also define the characteristic time scales for knots of transverse
size $D =(10^{16}~{\rm cm}) D_{16}$ impacted by internal shocks of velocity
$V_s$ and traversing a medium of density $n_{\rm e} = (10^3~{\rm cm}^{-3}) n_{3}$.  As
noted above, the internal shocks range from $100 - 500$ \kms, and the observed
knot sizes are typically $0\farcs1 - 0\farcs3$.  At the 3.4 kpc distance
estimated for Cas~A, an angular size of $0\farcs1$ corresponds to a linear size
of $5 \times 10^{15}$ cm.

The dynamical cloud crossing and gas cooling times are then:
\begin{eqnarray}
   t_{\rm cc}     &\equiv&  D/V_s \approx (30~{\rm yr}) D_{16} V_{100}^{-1}   \nonumber \\
   t_{\rm cool} &\equiv&  \frac {3kT/2} {n \Lambda(T)} \approx (100~{\rm yr}) 
         T_6^{1.7} n_3^{-1} Z^{-1}  \; .
\end{eqnarray}
Here, $Z$ is the plasma metallicity in units of solar values, and the radiative
cooling function $\Lambda(T) \approx (10^{-22}$~erg~cm$^3$~s$^{-1})T_6^{-0.7} \, (Z/Z_{\odot})$.
We define the temperature $T = (10^6~{\rm K}) T_6$, where the cooling
expression is valid over the range $0.3 < T_6 < 10$.  At temperatures
$1 < T_6 < 10$, the grain sputtering time is fairly constant \citep{DA07},
\begin{equation}
   t_{\rm sp} \approx (100~{\rm yr}) n_3^{-1} \left( \frac {a}{0.1~\mu{\rm m}} \right) \;  
\end{equation}
scaled to grains of radius $a \approx 0.1~\mu$m.   Thus, for temperatures $T_6 \approx 1$
and plasma densities $n \approx 10^3$~cm$^{-3}$, the grain sputtering time and
plasma cooling time are nearly equal and both are comparable to the cloud-crossing
time. \\

\subsection{Knot Flux Variability }

The basic idea behind emission variations in SN ejecta is that decade-long
variability like that seen in the remnant's main shell results from $\sim$100
km s$^{-1}$ shock-wave passage through ejecta knots with characteristic sizes
around $10^{16}$ cm.  Smaller knots
like those present in the remnant's outskirts and described in this paper
allow for faster variability with timescales at or below a few years.

Understanding ejecta knot brightness variability involves both dynamical and
radiative processes in the hot post-shock gas.  As described by Equations (1)
and (2),  the characteristic time scales for shock-crossing and radiative cooling
of post-shock gas are similar, ranging from 30--100 yr for densities $n_{e}
\approx 10^3$ cm$^{-3}$, knot sizes $D \approx 10^{16}$ cm, and solar
metallicities $Z \approx Z_{\odot}$.   Thus the slow, decades-long main shell
brightness variations reported by  \citet{kvdb76} and \citet{vdbk85} probably
reflect the passage of shocks through large ejecta knots of angular size
$\simeq1''$ ($5 \times 10^{16}$ cm). 

Shorter-term ($t \lesssim 1$~yr) variations like those seen in the outer knots
involve processes on much smaller scales, as these knots are compressed
and fragmented by instabilities arising from the shock passage.  Such processes
can be seen in the simulations of \citet{Mell02}, \citet{Raga07} and
\citet{Silvia10} which show that a shocked cloud with a relatively homogeneous
density structure will dissipated into a cluster of much smaller knots. 

Ejecta knots will, if $\lesssim 10^{15}$ cm in size ($\lesssim 0\farcs02$), 
exhibit optical variability on timescales around 1 yr or less. Thus, the 
observed rapid shock-induced flux variations on timescales of less than one
year suggests significant and dense knot substructures below the
diffraction resolution limit of {\sl HST}. However, we will show below in
$\S$4.3 indications from the lengths of knot emission tails that knot
variability is unlikely to be shorter than a few months.

A secondary but important ingredient for creating rapid variability is a highly
inhomogeneous interstellar or circumstellar medium. From a three year study of
infrared light echoes around Cas A, \citet{Kim08} find that the remnant's local
ISM is highly inhomogeneous, consisting of sheets and filaments on a scale of a
parsec or less. 

However, from our survey of outer knot flux variability, the remnant's
circumstellar environment appears to have structures on even smaller scales.
For example, the observed flickering of ejecta knots on times of months
indicates interactions with small-scale CSM features, as suggested by the rapid
flickering of the eastern limb knots C and D (Fig.\ 6).  In that case, the
invisible CSM cloud or sheet which knots C and D sequentially run into cannot
be much larger than $0\farcs5$ or about $0.01$ pc.

It is important to note that our investigation into rapid ejecta knot flux
variations was limited by the shortest time interval between images, namely the
nine-month separation of the 2004 ACS images.  Since it is unlikely that these
images captured a knot either at its maximum or its minimum flux brightness, the
optical flux variations illustrated in Figures $1 - 6$ and tabulated in Tables
1 and 2 may not necessarily represent the shortest times by which knot
substantially brighten or fade.  For example, an ejecta knot could have
brightened earlier than March 2004 and then begun to fade during our nine-month
observing period, meaning that our measured flux variation represents only a
minimum change in flux.  Similarly, if a knot brightened slowly, then our
observed flux change is again a minimum value.  Larger flux changes over longer
time scales are, in fact, suggested by the sudden appearance of knots in both
the jet and eastern limb regions.

\subsection{Knot Deceleration}

A knot will undergo deceleration both from the direct interaction with local
gas as well as from the internal shock driven into the knot that gives rise to
the optical emission observed \citep{Jones94}.  
If individual ejecta fragments are treated as dense undistorted knots, then their 
deceleration due to drag from its interaction with the ambient medium depends
on the knot's velocity and mass, the density of the local medium, and the
cross-sectional area of the knot's bow shock, which for hypersonic conditions is
approximately equal to the knot itself \citep{Chevalier75,Hamilton85,Jones94}.

The time scale for knot deceleration (drag) is given by $\tau_{drag}$ $\sim$
$\chi R_{k}$/$V_{k}$ where $\chi$ again is the density contrast between the
knot and the ambient medium (i.e.,  $\rho_{k}$/$\rho_{a}$), $R_{k}$ is the
knot's radius, and $V_{k}$ is the knot's velocity \citep{Jones94}.  Based on
ACS imaging data, typical outer knots have velocities of 10,000 km s$^{-1}$ and
diameters $\simeq 0\farcs1$ to  $\simeq 0\farcs3 $ corresponding to $\simeq 1
\times 10^{16}$ cm at 3.4 kpc \citep{HF08}.  High-velocity outlying [S~II]
emitting knots exhibit electron densities lie between $2000 - 16,000$ cm$^{-3}$ with
typical values between $4000 - 10,000$ cm$^{-3}$ \citep{Fes96,Fesen01}. 

Choosing  $\chi$ = 10$^{4}$ and an outer ejecta knot velocity of 10,000 km
s$^{-1}$ leads to $\tau_{drag}$ $\sim$ 2000 yr, suggesting that outer knot
deceleration due to drag may be fairly small.  However, if the knot fragments
into smaller knots as some models suggest \citep{Raga07,Silvia10}, smaller
knots will decelerate more rapidly owing to their larger surface area-to-mass
ratios. A significant deceleration of cloud fragments would be consistent
with observed lumps in the ablation tails of some ejecta knots like that shown
in Figures 8 and 9. 

\subsection{Knot Ablation Tails}

Cloud-ISM interaction models suggest knot disruption might also occur due to
Rayleigh-Taylor and Kelvin-Helmholtz instabilities resulting in both mass
ablation tails as we observed in some knots and the generation of smaller, dense
knot fragments \citep{Klein94,Jones94,Cid96,Raga07,Silvia10}.  The timescale
for initial knot breakup under these conditions, $\tau_{break}$, is uncertain but
is likely to be a few cloud crushing times or
$\tau_{break}$ $\sim$ 4 $\chi^{1/2} R_{k}$/$V_{k}$.

Model calculations by \citet{Pol04b} for a dense knot moving at 235 km s$^{-1}$
in a relatively dense medium but of similar size to the outer ejecta seen in
Cas A showed that deceleration of some of a knot's mass due to fragmentation of
the knot will result in a Hubble-like flow of ablated and stripped material
flowing off the knot.  The fact that most trailing emission tails we observed in
Cas A are relatively bright and sometimes clumpy suggest significant mass
stripping and partial knot break-up.  A possible example of knot
fragmentation is Knot 9 (see Fig.\ 8) where the knot appears noticeably
elongated in its direction of motion with a possibly associated fainter knot
some $0\farcs6$ behind which itself has some trailing emission.

The length of an observed ablation tail is limited by the decline in emission
measure ($EM = n_e^2 \, L$) as the hot post-shock gas expands and cools.
Because this stripped material may form a fairly a continuous medium of gas
with similar velocities and densities detached from the main knot, tail
material will not experience a strong internal shock.  Thus, trailing material
should fade in brightness in a timescale roughly equal to the material's
postshock cooling timescale. This can be estimated from the previous sections
regarding knot variability and these data suggest a timescale for substantial
fading on the order of several months.  Such a timescale is consistent with the
observed length of knot emission tails. For example, proper motions of Cas A's
outer knots is $\sim$ $0\farcs4 - 0\farcs9$ yr$^{-1}$ \citep{HF08} meaning that
a typical outer knot will move an angular distance roughly $0.2'' - 0.5''$ in a
timespan of $\sim0.5$ yr during which time its trailing ablated material cools
and hence fades. This agrees well with what is observed. 

Presumably all optical visible ejecta knots undergo some degree of mass loss
due to shock instabilities on the front and sides of the ejecta knot. However,
only a relatively small percentage ($\leq$ 5\%) of outer ejecta appear to have
visible ablation tails on broadband {\sl HST} ACS images. \citet{Pol04a} found
that knot fragmentation was key in producing significant wake emission.  To
investigate the formation of visible emission tails, we ran a series of model
calculations similar to those presented in \citet{Silvia10} but where a knot of
initial density $1000$ cm$^{-3}$ and temperature of $10^{4}$ K and an oxygen
abundance 100 times solar passed through a medium of density of $1$ cm$^{-3}$
with a velocity of 10,000 km s$^{-1}$. Under such conditions, a substantial tail
in terms of emission measure was formed. Our simulations also showed that with
$\chi \sim100$, the knot was disrupted too quickly to show a strong
head--tail structure, whereas if $\chi \gg 10^{4}$ then too little material was
ablated in a trailing wake to be readily visible.

Prior model simulations of small knots have suggested that the inevitable
instabilities associated with shock heating are likely to have short lifetimes
on the order of $t_{cc}$ \citep{Jones94}, which for a $0\farcs2$ diameter knot
with an internal shock velocity $\simeq$ 100 km s$^{-1}$ is about 30 yr. This
timeframe is consistent with observations of the brighter of Cas A's outer
ejecta knots which have been bright enough to be optically visible for several
decades and some of these show emission tails in {\sl HST} images.  One example
is the trailing emission Knot 9 shown in Figure 8 (referred to as Knot $10$ in
\citealt{Fesen01}) which is visible on a deep red 1976 Palomar 5~m photographic
plate (PH~7252vb: 098-04 emulsion + RG645 filter; \citealt{vdbk83}.  Thus this
knot, which exhibits very weak H$\alpha$ emission but strong [\ion{N}{2}]
emission lines, has been optically bright for nearly three decades
\citep{Fesen01}.

Finally, we note that the emission tails seen for Cas A's outer ejecta knots
appear morphologically different from the emission spikes or ``strings'' seen
emanating from the $\eta$ Carinae Homunculus nebula
\citep{Meaburn87,Meaburn96,Morse98,Weis99}. These long ($4''- 16''$) and thin
($0\farcs1 - 0\farcs25$) emission features are thought to be decelerated
ablated mass loss caused by the motion of $\sim 1000$ km s$^{-1}$ `bullets' of
ejected material moving through the star's surrounding circumstellar medium
\citep{Redman02,Currie03,Pol04b}. 

Unlike the trailing emission features seen in Cas A, however, the five $\eta$
Carinae strings are not perfectly straight, they seem to fade in brightness
with increasing distance from $\eta$ Carinae, and show no obvious optical
emission `head' or leading knot of emission. These properties are quite unlike
what is observed in the trailing ablation flows associated with Cas A's ejecta
knots. 

The lack of an obvious knot at the tip of any of these emission strings has
been attributed to very high postshock temperatures ($\gtrsim 10^{6}$ K)
initially present in the ablated material making them appear weak optically
\citep{Redman02,Pol04b}.  Interestingly, exactly the reverse is observed in Cas
A were we see emission directly attached to the hypersonic knots which have
expansion velocities an order of magnitude greater that seen in $\eta$ Carinae.  

We speculate that the emission strings seen in $\eta$ Carinae represent ablation
flows in which ejected knots are largely destroyed and dissolved by their passage
through the relatively dense circumstellar medium around $\eta$ Carinae. This
would account for both the fading of the strings with increasing distance and
the lack of any observable optical bullet head at the string tips.

\subsection{Ejecta Fragmentation}

Besides the noted clumpiness of some emission tails, there are additional
indicators for the disruption and fragmentation of ejecta knots, in particular
the break-up of relatively large ones. The evidence comes from the presence of
several unusually tight clusters of ejecta knots present in the remnant's NE
jet.  These clusters consist of several closely spaced knots of similar size
and brightness.  Four examples of such knot clusters present in the NE jet are
shown in the upper panels of Figure 11.  

\begin{figure*}
\includegraphics[width=\linewidth]{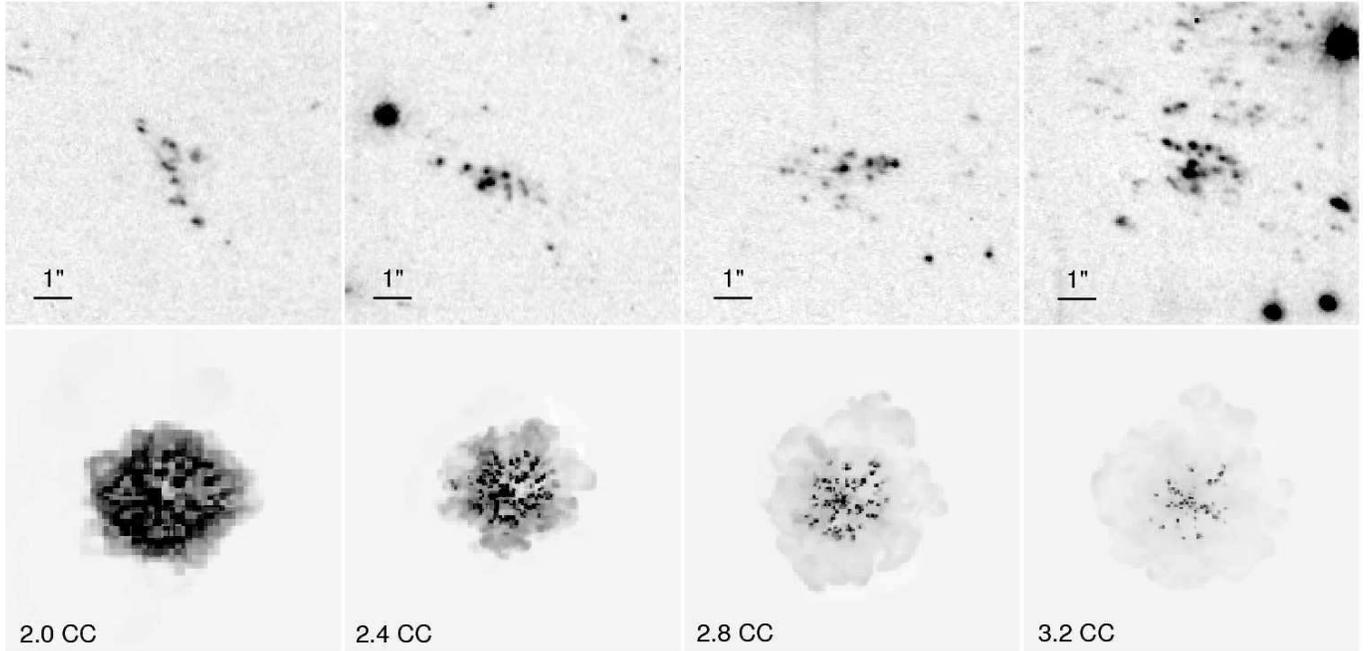}
\caption{ {\it{Top row:}} Enlargements of the December 2004 ACS F675W + F775W image
showing four ejecta knot clusters in the NE jet. 
{\it{Bottom Row:}} Formation of knot clusters due to the fragmentation of 
a large ejecta clump after 2.0, 2.4, 2.8 and 3.2 cloud crushing times \citep{Silvia10}. }
\label{fig:knot_cluster_n_models}
\end{figure*}

Morphologically, these tight knot clusters appear similar to the results of
shock models which show the fragmentation of a large ejecta knot into a tight
cluster of small, denser knots \citep{Mell02,Raga07,Silvia10}.  To illustrate
this, we show in the lower panel of Figure 11 model results from
\citet{Silvia10} for the evolution of ejecta into a knot cluster over a
timespan of a few cloud-crushing time scales.  Although the jet knot clusters
are aspherical and hence look different morphologically from the spherical knot
clusters seen in the model results, the exact structure knot clusters will of
course depend upon the initial size and structure of the ejecta cloud prior to
shock passage and is unlikely to be spherical.

Other examples of small clusters of ejecta knots are two ejecta knot clusters
situated along the remnant's northern periphery. Although initially thought to
be two single knots from inspection of ground-based images (knots 15 and 15A;
see \citealt{Fesen01}), {\sl HST} images resolve them into two tight groups of
knots each consisting of  a dozen or so individual knots (see on-line Figure
Sets 2 \& 3 for Region H; \citealt{HF08}).  The outlying Knot 15 (also known as
Knot 91, \citealt{kvdb76}) is of special interest for estimating lifetimes of
ejecta fragments formed from the fragmentation of a larger knot. Although
recently showing signs of fading, this group of ejecta knots has remained
optically visible for nearly 60 years, being easily detected on one of the
first optical images taken of the remnant, namely a November 1951 Palomar 5~m
plate (PH 563b; \citealt{vdbD70}).  

\section{Conclusions}

We report results of a survey on outlying debris knots around
the young core-collapse SN remnant Cassiopeia A using {\sl Hubble Space
Telescope} images. These images reveal ejecta knot flickering, ablation
emission tails, and knot fragmentation which are phenomena associated with the
initial stages of the gradual merger of SN ejecta with, and the enrichment of,
the surrounding interstellar medium. 

Our major findings and conclusions are:

1) Substantial changes in the optical flux were seen in a significant
percentage of the remnant's outer ejecta knots including the NE jet and eastern
limb regions.  In general, we detected no pattern in terms of brightness
changes between three filters to suggest systematic changes in the line
emissions between knot brightening or fading that might suggest a change of a
knot's dominant ionization state.

2) A variety of light curves was found with some knots going up and down
in brightness (`flickering') in times as short as nine months.  Flux changes
typically ranged from 50\% to 800\%.  While large brightness changes were only
seen in a minority of jet emission knots ($<$25\%), smaller flux changes were
apparent in a much larger percentage of knots approaching 50\%.

3) The observed flickering of ejecta knots on times of months indicates knot
scale lengths $\lesssim 10^{15}$ cm (i.e., $\lesssim 0\farcs02$) along with  a
highly inhomogeneous interstellar or circumstellar medium.  Knot interaction
with small scale CSM features is suggested by the rapid flickering of two 
eastern limb knots, C and D (see Fig.\ 6).  In that case, the invisible CSM cloud or
sheet which knots C and D sequentially interacted with, leading to their
sequential brightening, cannot be much larger than $0\farcs5$ or about $0.01$
pc.

4) A small percentage ($<$5\%) of knots were found to exhibit trailing emissions
$0\farcs2 - 0\farcs7$ in length aligned along a knot's direction of
motion suggestive of mass ablation tails due to high-velocity interaction
with local CSM and ISM.  These emission tails are bright relative to the knot
suggestive of considerable ablated mass loss and could be either smooth or
lumpy in intensity away from the knot's head.  Because these linear knot + tail
features are relatively short and brighter near the knot, they are unlike the
linear emission ``strings'' seen in $\eta$ Carinae.  
  
5) We identified several tight clusters of ejecta
knots which closely resemble fragmentation models for larger SN ejecta knots 
caused by a high-velocity interaction with a lower density ambient medium.
Archival images of Cas A suggest that these knot clusters can survive and
remain bright for several decades.

\acknowledgements

This work was supported by NASA through grants GO-8281, GO-9238, GO-9890, and
GO-10286 from the Space Telescope Science Institute (STScI), which is
operated by the Association of Universities for Research in Astronomy.  JMS
acknowledges support at the University of Colorado through theoretical grants
from STScI (AR-11774.01-A), NASA (NNX07-AG77G) and NSF (AST07-07474).

\clearpage
\newpage


\end{document}